\documentclass[twocolumn,prb]{revtex4}
\usepackage{graphicx}
\usepackage{dcolumn}
\usepackage{bm}
\usepackage{amsmath}

\begin{document}

\preprint{}
\title{Theory of pairing symmetry in  Fulde-Ferrell-Larkin-Ovchinnikov vortex state and vortex lattice}
\author{Takehito Yokoyama$^1$, Masanori Ichioka$^2$, and Yukio Tanaka$^1$}
\affiliation{$^1$ Department of Applied Physics, Nagoya University, Nagoya, 464-8603, Japan
\\
$^2$ Department of Physics, Okayama University, Okayama 700-8530, Japan}
\date{\today}

\begin{abstract}
We investigate pairing symmetry in the  Fulde-Ferrell-Larkin-Ovchinnikov (FFLO) vortex state and vortex lattice, and explain the electronic structure in these states in terms of pairing symmetry. 
We show analytically 
that at the intersection point of FFLO nodal plane and vortex line, only even frequency pairing is present if the Zeeman splitting is negligibly small. With increasing Zeeman splitting, odd frequency pairing emerges there. This makes it possible to interpret the
gap structure of the density of states at the intersection point as a manifestation of the even frequency pairing. 
In the vortex lattice, we find that only odd frequency pairing is present at the core centers, while at the midpoint of the vortex lines, only even frequency pairing exists. Thus, the odd and even frequency pairings also form the lattice in the vortex lattice state.

\end{abstract}

\maketitle

%
%


\section{Introduction}

The study of the mixed state in type-II superconductors has
a long history and revealed a wide variety of physical
phenomena.\cite{Abrikosov}
In the clean limit,
low-energy bound states, dubbed Andreev bound states, are formed in
the vortex core due to the spatial profile of the superconducting
pair potential. \cite{Caroli et al,Hess} The bound states manifest themselves as an enhancement of
zero-energy quasiparticle density of states (DOS) in the
core, observable as a zero-bias conductance peak by scanning
tunneling microscope.\cite{Hess,Fischer} Recently, the Andreev bound states have been revisited from the viewpoint of \textit{the odd frequency pairing. }\cite{Yokoyamavortex} 
  
Generally, superconducting pairing is classified into
even-frequency or odd-frequency state according to a symmetry with
respect to time. Due to the Fermi statistics, even-frequency
superconducntors belong to the symmetry class of spin-singlet even-parity
 or spin-triplet odd-parity pairing state, while odd-frequency
superconductors belong to the spin-singlet odd-parity or
spin-triplet even-parity pairing state.

Although the possibility of the odd-frequency pairing state in various kinds
of uniform systems was discussed in the literature,
\cite{Berezinskii,Fuseya} its realization in
bulk materials is still controversial.
On the other hand, the
realization of the odd-frequency pairing state has recently been pointed out in inhomogeneous even-frequency superconducting systems such as ferromagnet/superconductor junctions,\cite{Efetov1} normal metal/superconductor interface,\cite{Tanaka3,Linder} Abrikosov vortex\cite{Yokoyamavortex}  or a diffusive normal metal attached to a spin-triplet superconductor.\cite{Tanaka2006}

In Ref. \onlinecite{Yokoyamavortex}, pairing symmetry in an isolated vortex is elucidated. It is found that at the center of the core, purely odd frequency pairing state exists. Since single vortex is considered in  Ref. \onlinecite{Yokoyamavortex}, this theory is applicable to the low field regime. On the other hand, at high magnetic fields, the distance between vortices becomes short, and hence the overlap effect of the vortex cores cannot be neglected. 
 In this situation, the local DOS
around a vortex core is expected to break cylindrical symmetry
and show sixfold symmetric structure when the vortex
lattice forms a triangular lattice.\cite{Hayashi,ichiokaS} Also, at high field regime,  Fulde-Ferrell-Larkin-Ovchinnikov (FFLO) vortex state may be realized under some conditions.\cite{Matsuda} These states are also inhomogeneous systems, which may provide rich structure of superconducting correlation, in particular odd frequency pairing. 

The FFLO state~\cite{ff,lo} is realized by Cooper pairs with non-zero center of mass momentum when the Fermi surfaces for up-spin and down-spin electron bands are largely split by the Zeeman effect, 
thus inducing the spatial modulation of the pair 
potential.~\cite{machida,tachiki,shimahara,klein,buzdin,adachi,ikeda,Vorontsov} 
Several experiments support the realization of the FFLO state in a high field phase of a quasi-two dimensional (Q2D) heavy Fermion superconductor 
${\rm CeCoIn_5}$.~\cite{bianchi,radovan,watanabe,capan,martin,kakuyanagi,kumagai} 
There, it is supposed that the FFLO nodal planes of the pair potential run 
perpendicular to the vortex lines. 

In general, we have to include the vortex structure 
in addition to the FFLO modulation, because the FFLO state appears at high fields in the mixed states.\cite{buzdin,adachi,ikeda} 
The vortex and FFLO nodal plane structures 
in the FFLO state were studied in Refs.~\onlinecite{mizushima2,Ichioka}.  
It is found that the topological structure of the pair potential 
plays important roles to determine the electronic structures 
in the FFLO vortex state. 
The pair potential has $2\pi$-phase winding around the vortex line, and 
$\pi$-phase shift at the nodal plane of the FFLO modulation. 
These topologies of the pair potential structure affect the distribution 
of paramagnetic moment and low energy electronic states  
inside the superconducting gap. 

Another aspect of the FFLO state is the parity mixing. In the FFLO state, in addition to the breakdown of the translational symmetry, Zeeman field breaks the SU(2) symmetry in spin space. As a consequence, singlet-triplet mixing state emerges as a stable phase in the FFLO state. \cite{Matsuo,Yokoyama,Yanase,Aizawa}
Therefore, quite gerenally, in the FFLO state mixture of even-odd frequency and singlet-triplet pairings is expected to emerge. 

In this paper, based on the quasiclassical theory of superconductivity, we investigate pairing symmetry in the FFLO vortex and vortex lattice, and explain the electronic structure in these states in terms of pairing symmetry. 
We show analytically 
that at the intersection point of FFLO nodal plane and vortex line, only even frequency pairing is present if the Zeeman splitting is negligibly small. With increasing Zeeman splitting, odd frequency pairing  emerges there. This makes it possible to interpret the
gap structure of the density of states at the intersection point as a manifestation of the even frequency pairing. 
In the vortex lattice, at the core centers, only odd frequency pairing is present
while at the midpoint of the vortex lines, only even frequency pairing
appears irrespective of the energy.
Therefore, the odd and even frequency pairings also form the lattice in the vortex lattice state.

The organization of the paper is as follows. In Sec. II, we study the FFLO vortex state, and show the electric structure and superconducting correlation of the system. Sec. III is devoted to the study of the vortex lattice. 
The summary is given in Sec. IV.

\section{FFLO vortex}
\subsection{Formulation}
The electronic structure of the vortex core in an inhomogeneous 
clean superconductor can be described by the quasiclassical Eilenberger
equations \cite{Eilenberger,Larkin}
%
based on the Riccati parametrization\cite{SchopohlMaki}.  
Along a trajectory
${\bm r}(x') = {\bm r}_0 + x' \; \hat{\bm {v}}_F $ with unit vector
$ \hat{\bm v}_F$  parallel to Fermi velocity $\bm {v}_F$, the Eilenberger equations
are generally represented in 4$\times4$ matrix form\cite{Eschrig}, which, in terms of the Ricatti parameters $\hat a$ and $\hat b$, reduces to:\cite{SchopohlMaki}
\begin{eqnarray}
v_F \partial _{x'} \hat a + 2\omega_n \hat a + \hat a \hat \Delta ^\dag  \hat a - \hat \Delta  + i \Sigma  \hat a - i\hat a \tilde \Sigma   = 0 \\ 
v_F \partial _{x'} \hat b - 2\omega_n \hat b - \hat b \hat \Delta \hat b + \hat \Delta ^\dag   - i\hat b\Sigma   + i\tilde \Sigma  \hat b = 0
\end{eqnarray}
with the self energy of the Zeeman splitting
$\Sigma   = \tilde \Sigma   = \mu_B H \sigma _z$ and  Pauli matrix $\sigma _i (i=x,y,z)$.
Here, we have defined 2$\times2$ matrix $\hat a$ and $\hat b$ in spin space via  the 4$\times4$ Green's functions $\check{g} $:
\begin{eqnarray}
 \check{g} = \left( {\begin{array}{*{20}c}
   {\hat g} & {\hat f}  \\
   {\hat f^\dag  } & { - \hat g}  \\
\end{array}} \right) \nonumber \\ 
  \equiv  - \left( {\begin{array}{*{20}c}
   {(1 + \hat a\hat b)^{ - 1} } & 0  \\
   0 & {(1 + \hat b\hat a)^{ - 1} }  \\
\end{array}} \right)\left( {\begin{array}{*{20}c}
   {1 - \hat a\hat b} & {2i\hat a}  \\
   { - 2i\hat b} & { - (1 - \hat b\hat a)}  \\
\end{array}} \right) .
\end{eqnarray}

There are two possible spatial modulation 
of the pair potential $\Delta$ in the FFLO states. 
One is the Fulde-Ferrell (FF) state~\cite{ff} with phase modulation 
such as $\Delta \propto {\rm e}^{{\rm i}Qz}$, 
where $Q$ is the modulation vector 
of the FFLO states.  
The other is the Larkin-Ovchinnikov (LO) state~\cite{lo} 
with the amplitude modulation  such as $\Delta \propto \sin Qz$, 
where the the pair potential shows the periodic sign change, 
and $\Delta=0$ at nodal planes. 
We discuss the case of the LO states in this paper, 
since some experimental~\cite{watanabe,capan,kakuyanagi} 
and theoretical~\cite{buzdin,ikeda} works 
support the LO state (at least in low temperature region)
for  the FFLO states in  ${\rm CeCoIn_5}$. 

\begin{figure}[htb]
\begin{center}
\scalebox{0.4}{
\includegraphics[width=17.0cm,clip]{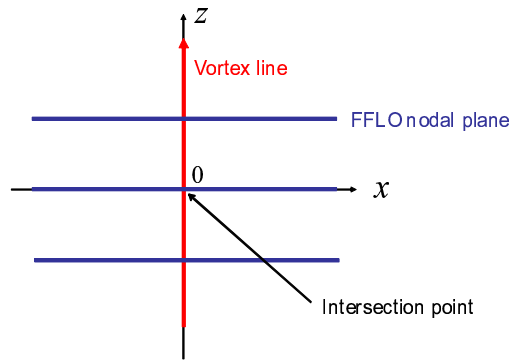}}
\end{center}
\caption{(color online) Configurations of the vortex line and the FFLO nodal
planes in the $x-z$ plane. } \label{f0}
\end{figure}

When we consider vortex structure in the LO state, 
there are two possible choices of the configuration 
for the vortex lines and the FFLO modulation: the modulation vector of the FFLO state is parallel~\cite{tachiki} 
or perpendicular~\cite{shimahara,klein} to the applied magnetic field. 
In this paper, we study the former case by the quasiclassical 
theory as shown in Fig. \ref{f0}.~\cite{tachiki,klein,klein87,ichiokaS,ichiokaD,ichiokaMgB2} 

As a model of Fermi surface in ${\rm CeCoIn_5}$, 
we use a Q2D Fermi surface with rippled cylinder-shape, 
and the Fermi velocity is given by 
${\bf v}_{\rm F}=(v_a,v_b,v_c) \propto 
( \cos\theta, \sin\theta, \tilde{v}_z \sin k_c)$ 
at the Fermi surface 
${\bf k}_{\rm F}=(k_a,k_b,k_c)
\propto(k_{\rm F0}\cos\theta,k_{\rm F0}\sin\theta,k_c)$ 
with $0 \le \theta \le 2 \pi$ and $-\pi \le k_c \le \pi$.~\cite{ichiokaMgB2} 
In our calculation we set $\tilde{v}_z =0.5$, so that the anisotropy ratio 
$\gamma=\xi_c/\xi_{ab} \sim  
\langle v_c^2 \rangle_{\bf k}^{1/2} /\langle v_a^2 \rangle_{\bf k}^{1/2} 
\sim 0.5$.  Here, $\bar{v}_{\rm F}=\langle v_{\rm F}^2 \rangle_{\bf k}^{1/2}$ 
is an averaged Fermi velocity on the Fermi surface. 
$\langle \cdots \rangle_{\bf k}$ indicates the Fermi surface average. 
A magnetic field is applied along the $a$ axis direction in our calculation. 
Thus, the coordinate $(x,y,z)$ for the vortex structure corresponds to 
$(b,c,a)$ of the crystal coordinate.

We consider $d_{x^2-y^2}$-wave FFLO vortex state, and choose the following form of the pair potential:
\begin{eqnarray}
\hat \Delta ({\bf{r}},\theta ) = \Delta _0 \cos 2\theta \tanh \left( {\frac{{\sqrt {x^2  + y^2 } }}{\xi }} \right)\exp (i\varphi )\sin Qz\sigma _y 
\end{eqnarray}
with  $\exp(i\varphi)=(x + iy)/\sqrt {x^2  + y^2 }$.
Here, we introduce the coherence length $\xi = \hbar \bar{v}_{\rm F}/\Delta_0$,  the vortex line is situated at $x=y=0$, and $\exp(i\varphi)$ is
the phase factor which originates from the vortex. 

For the calculation of the local
DOS normalized by its value in the normal state,
 the quasiclassical propagator has to be integrated over $\theta$ and $k_z$ which define the direction of the Fermi velocity.
The normalized local DOS in terms of functions $\hat a$ and $\hat b$ is given by
\begin{widetext}
\begin{equation}
N({\bf{r}}_0 ,E) = \frac{1}{2}{\rm{Tr}} \int_{ - \pi }^\pi  {\frac{{dk_z }}{{2\pi }}} \int_0^{2\pi } {\frac{{d\theta }}{{2\pi }}} {\mathop{\rm Re}\nolimits} \left[ {(1 + \hat a\hat b)^{ - 1} (1 - \hat a\hat b)} \right]_{\omega _n  \to E + i\delta } \label{angularaverage}
\end{equation}
\end{widetext}
where $E$ denotes the quasiparticle energy with respect to the Fermi
level and $\delta$ is an effective scattering parameter. In numerical calculations throughout this paper, we will fix this value as
$\delta=0.05\Delta_0$.

We obtain singlet (triplet) pair amplitude $f_{s(t)}$ as
\begin{eqnarray}
f_{s(t)}  = \left\{ {f_{}^{ \uparrow , \downarrow }  - ( + )f_{}^{ \downarrow , \uparrow } } \right\}/2.
\end{eqnarray}
Its even (odd) frequency component, $f_{even(odd)}^s$ and $f_{even(odd)}^t$, is obtained as 
\begin{eqnarray}
 f_{even(odd)}^s  = \left\{ {f_s (\theta ,k_z ) + ( - )f_s (\theta  + \pi , - k_z )} \right\}/2, \\ 
 f_{even(odd)}^t  = \left\{ {f_t (\theta ,k_z ) - ( + )f_t (\theta  + \pi , - k_z )} \right\}/2 .
\end{eqnarray}
Note that due to the Fermi statistics, for singlet pairing, even (odd) parity state should be even (odd) frequency pairing, while  for triplet pairing, even (odd) parity state should be odd (even) frequency pairing.

Their average in momentum space is defined as
\begin{eqnarray}
\left\langle {f_{even(odd)}^A } \right\rangle ^2  = \frac{1}{{4\pi ^2 }}\int_{ - \pi }^\pi  {\int_0^{2\pi } {\left| {f_{even(odd)}^A } \right|^2 d\theta } dk_z } 
\end{eqnarray}
with $A=s, t$. 
The magnitude of the even (odd) frequency pairing can be repesented as 
\begin{eqnarray}
\left\langle {f_{even(odd)} } \right\rangle ^2  =\left\langle {f_{even(odd)}^s } \right\rangle ^2+\left\langle {f_{even(odd)}^t } \right\rangle ^2.
\end{eqnarray}




\subsection{Results}
In the FFLO vortex state, along the trajectory through the intersection
point of a vortex and a nodal plane, the gap function does not
change sign, because the phase shift is 2$\pi$ by summing $\pi$ due to the vortex and $\pi$ due to the nodal plane.\cite{mizushima2,Ichioka} Consequently, the DOS at the intersection point does not have the Andreev bound state which is seen in the conventional vortex state or FFLO nodal plane. 
Our analysis presented below makes it possible to understand these features of DOS from the viewpoint of the  symmetry of superconducting correlations. 

First, we discuss the general property of the symmetry in the FFLO vortex state.  Consider a trajectory passing through the intersection point of vortex line and FFLO nodal plane. 
By setting $x'=0$ at the intersection point, we obtain $\hat b(x',\omega _n ) = \sigma _y \hat a( - x', - \omega _n )^{ - 1} \sigma _y $ from the Eilenberger equations since the gap function does not change sign at $x'=0$ and $\hat \Delta(x')=\hat \Delta(-x')$.

 Thus, at the intersection point $x'=0$,  we have $\hat f(0,\omega _n ) = \sigma _y \hat f(0, - \omega _n )\sigma _y$, and hence $ f_s (0,\omega _n ) = f_s (0, - \omega _n )$ and $ f_t (0,\omega _n ) =  -  f_t (0, - \omega _n )$. This means that at the intersection point, only even-frequency spin-singlet even-parity and odd-frequency spin-triplet even-parity parings are allowed to exist [Note that even (odd) frequency singlet (triplet) pairing should have even parity in accordance with the Fermi statistics].  In particular, at $\mu_B H=0$ since spin is conserved, at the intersection point only even-frequency spin-singlet even-parity pairing exists. With increasing $\mu_B H$, odd-frequency spin-triplet even-parity  pairing emerges there.  In a similar way, we can show that at the center of the FFLO nodal plane (vortex core) without vortex (the FFLO modulation), $\hat f(0,\omega _n ) = -\sigma _y \hat f(0, - \omega _n )\sigma _y$ is satisfied since the gap function changes sign at $x'=0$ and $\hat \Delta(x')=-\hat \Delta(-x')$. Hence, only odd-frequency spin-singlet odd-parity and even-frequency spin-triplet odd-parity pairings are allowed there. In particular, at $\mu_B H$=0, only odd-frequency spin-singlet
odd-parity pairing is present. 
The suppression (enhancement) of the local DOS at zero energy is related to the presence of even (odd)-frequency
pairing.\cite{Tanaka3,Tanaka2006,Yokoyamavortex} It has been shown that the emergence of the odd-frequency pairing is a physical reason of zero energy peak of the local DOS inside the vortex core.\cite{Yokoyamavortex}

\begin{figure}[htb]
\begin{center}
\scalebox{0.4}{
\includegraphics[width=17.0cm,clip]{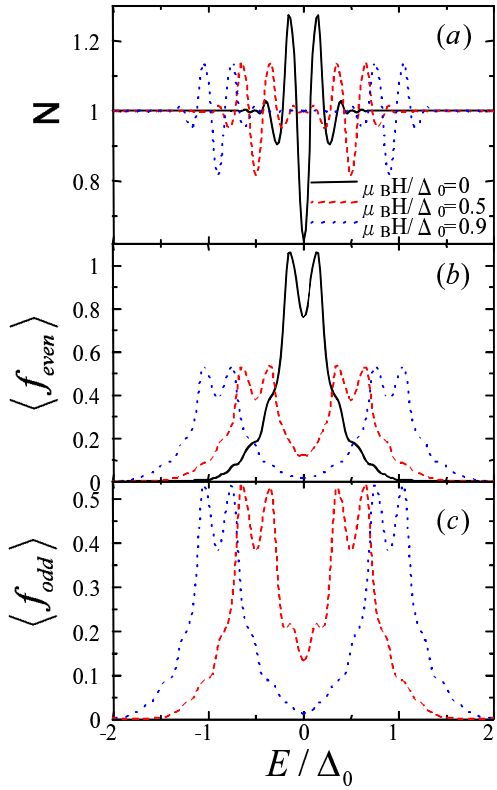}}
\end{center}
\caption{(color online) Normalized local DOS at the intersection point $(x,z)=(0,0)$.  } \label{f1}
\end{figure}

\begin{figure}[htb]
\begin{center}
\scalebox{0.4}{
\includegraphics[width=17.0cm,clip]{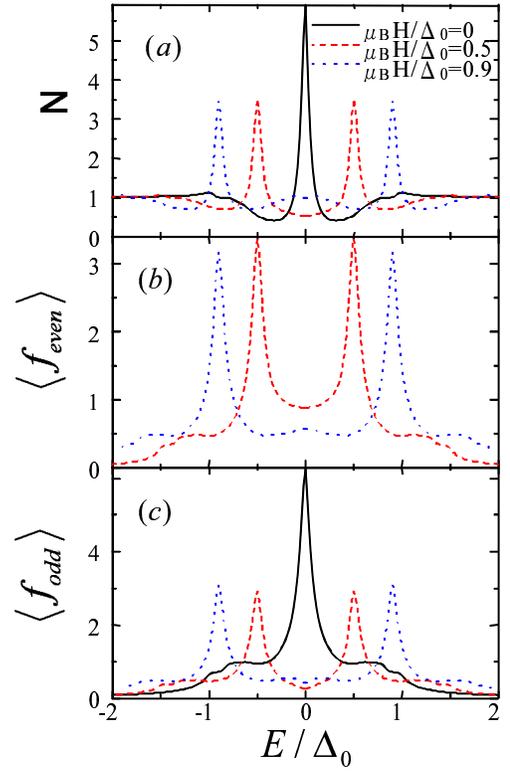}}
\end{center}
\caption{(color online) Normalized local DOS at the the center of the core  $(x,z)=(0,25 \xi)$.  } \label{f2}
\end{figure}

\begin{figure}[htb]
\begin{center}
\scalebox{0.4}{
\includegraphics[width=17.0cm,clip]{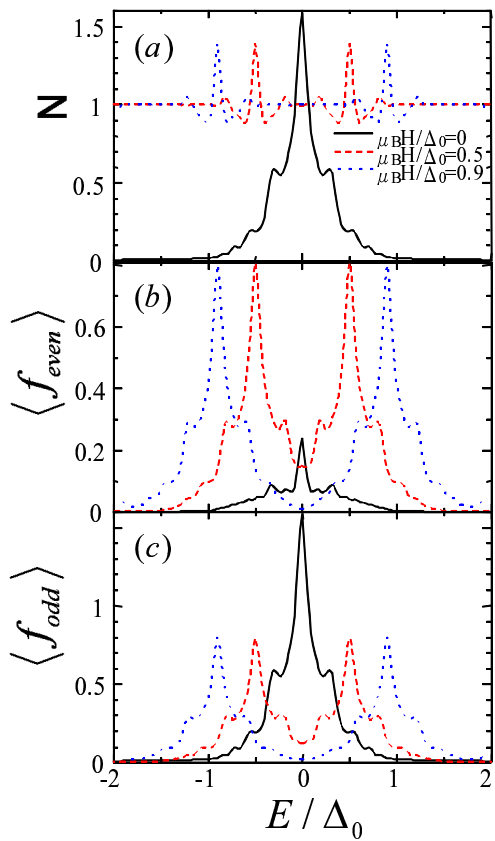}}
\end{center}
\caption{(color online) Normalized local DOS at the nodal plane $ (x,z)=(25\xi,0)$.   } \label{f3}
\end{figure}

\begin{figure}[htb]
\begin{center}
\scalebox{0.4}{
\includegraphics[width=17.0cm,clip]{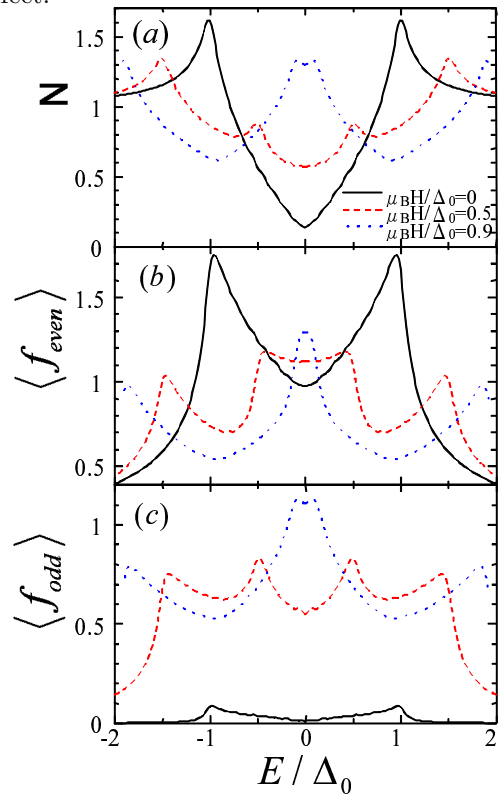}}
\end{center}
\caption{(color online) Normalized local DOS away from the nodal plane and the core $(x,z)=(25\xi,25 \xi)$.  } \label{f4}
\end{figure}

\begin{figure*}[htb]
\begin{center}
\scalebox{0.4}{
\includegraphics[width=31.0cm,clip]{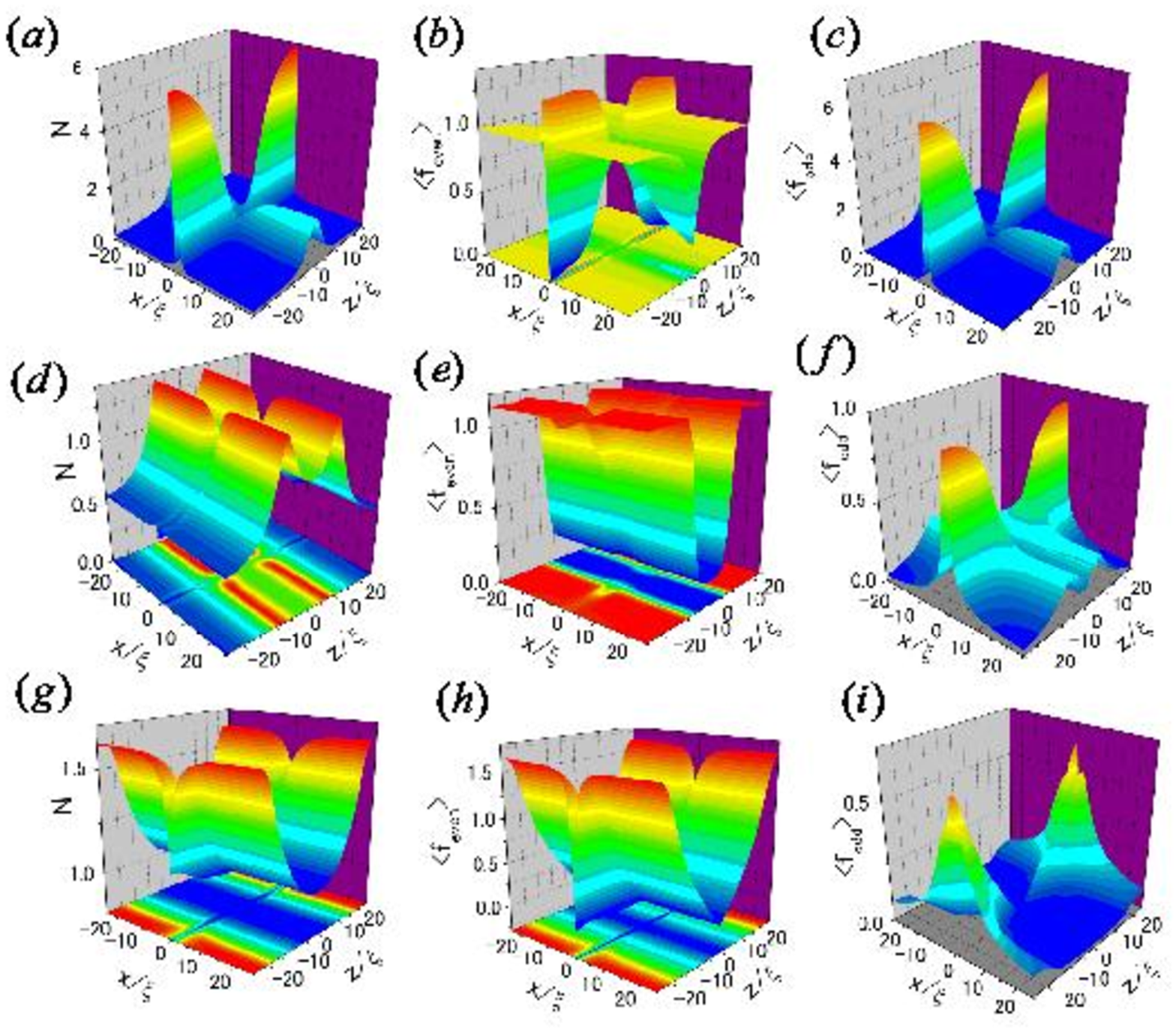}}
\end{center}
\caption{(color online) (a, d, g) normalized local DOS. (b, e, h) magnitude of even frequency pairing. (c, f, i) magnitude of odd frequency pairing. (a-c) $E=0$.  (d-f) $E=0.5 \Delta_0$.  (g-i) $E= \Delta_0$.  Here, we set $\mu_B H=0$. } \label{f5}
\end{figure*}

\begin{figure*}[htb]
\begin{center}
\scalebox{0.4}{
\includegraphics[width=31.0cm,clip]{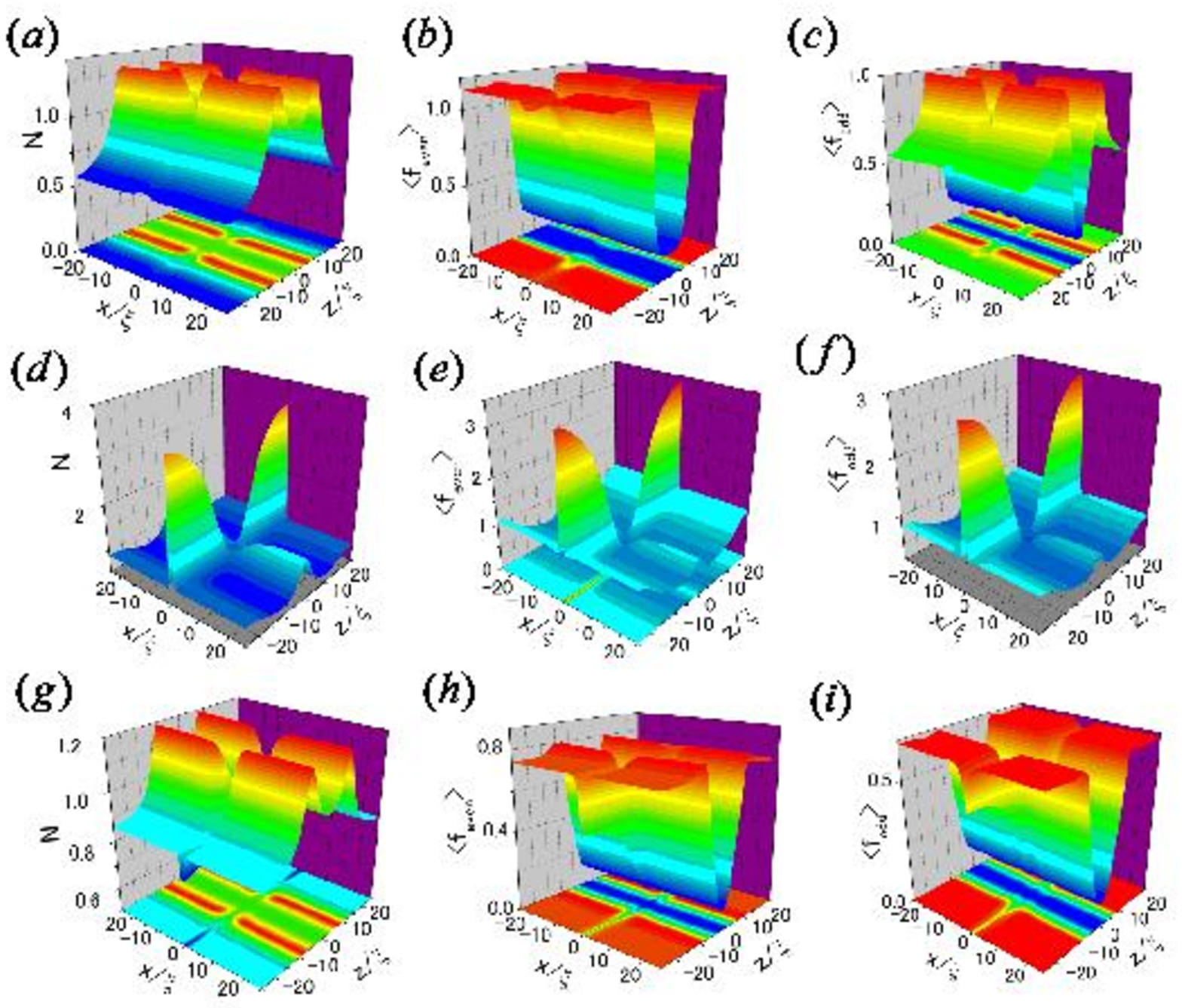}}
\end{center}
\caption{(color online) (a, d, g) normalized local DOS. (b, e, h) magnitude of even frequency pairing. (c, f, i) magnitude of odd frequency pairing. (a-c) $E=0$.  (d-f) $E=0.5 \Delta_0$.  (g-i) $E= \Delta_0$.  Here, we set $\mu_B H=0.5 \Delta_0$. } \label{f6}
\end{figure*}

\begin{figure}[htb]
\begin{center}
\scalebox{0.4}{
\includegraphics[width=22.0cm,clip]{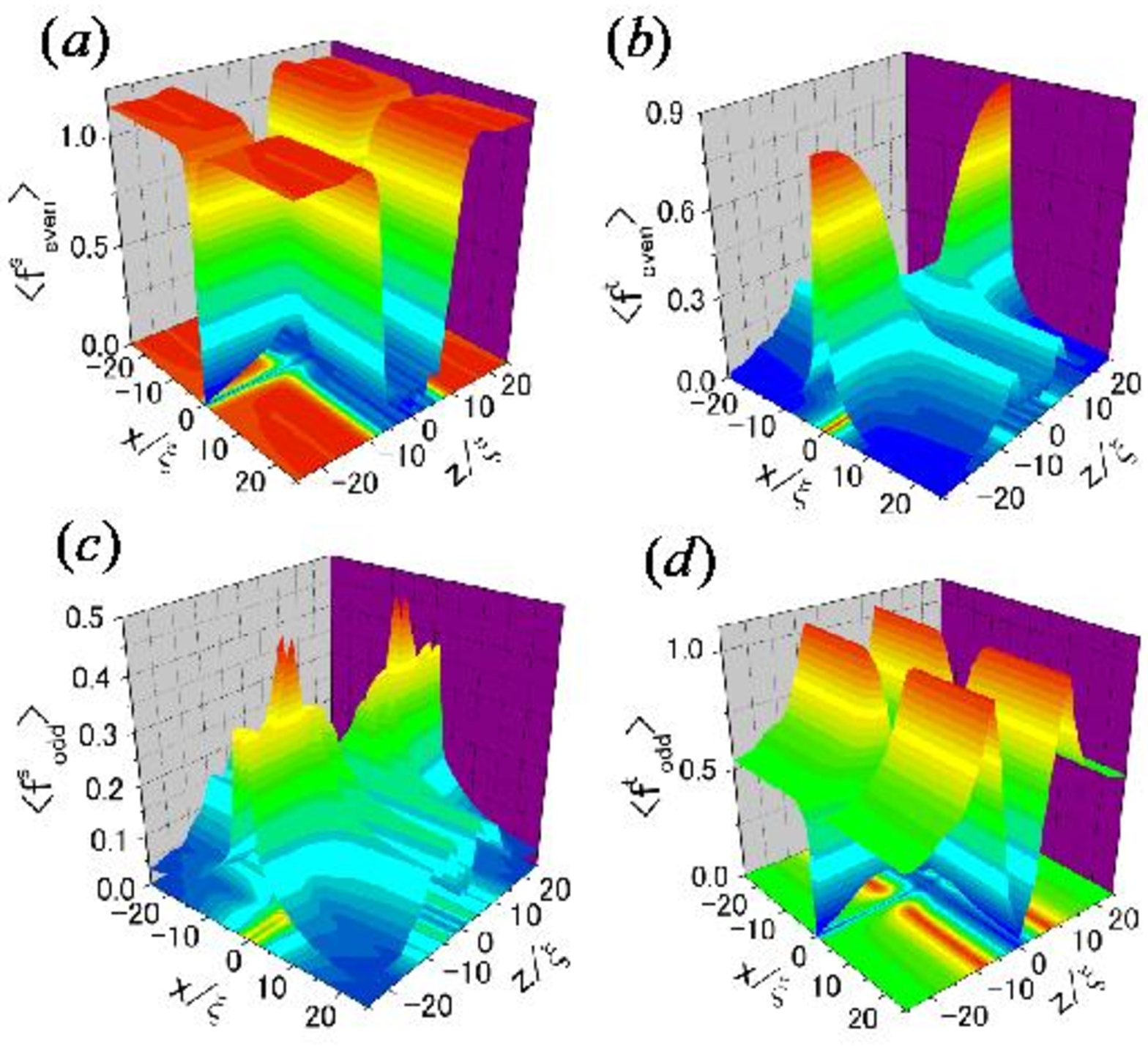}}
\end{center}
\caption{(color online) (a) magnitude of even frequency spin singlet even parity pairing. (b) magnitude of even frequency spin triplet odd parity pairing.
(c) magnitude of odd frequency spin singlet odd parity pairing.
(d) magnitude of odd frequency spin triplet even parity pairing.
Here, we set $E=0$ and $\mu_B H=0.5 \Delta_0$. } \label{f10}
\end{figure}


Next, we show the results of DOS and pair amplitudes by numerically solving the Eilenberger equation.
In the following we set $Q=2\pi/(100\xi)$ and $y=0$.

Figure \ref{f1} shows (a) the local DOS, (b) magnitude of even frequency pairing and (c) magnitude of odd frequency pairing as a function of the energy at the intersection point $(x,z)=(0,0)$.  At $\mu_B H=0$, local DOS shows a gap structure, which reflects that at the intersection point, only even frequency pairing exists at $\mu_B H=0$ as seen in Figs. \ref{f1} (b) and (c). 
With increasing $\mu_B H$, these structures are split and odd frequency pairing emerges there. As shown in the above discussion, at the intersection point, only even parity pairing is allowed to exist. Thus, even (odd) frequency pairing should be spin singlet (triplet).

Figure \ref{f2} displays (a) the local DOS, (b) magnitude of even frequency pairing and (c) magnitude of odd frequency pairing as a function of the energy at the center of the core, $(x,z)=(0,25 \xi)$. At $\mu_B H=0$, local DOS shows a peak structure, reflecting the presence of the odd frequency pairing for $\mu_B H=0$ as seen in Figs. \ref{f2} (b) and (c). 
As $\mu_B H$ increases, these structures are shifted and even frequency pairing emerges there. Note that here spin singlet (triplet)  state dominates spin triplet (singlet) state in even (odd) frequency pairing. This can be understood by considering the vortex without the FFLO nodal plane. In this case,  only odd parity pairing is allowed at the core center. Therefore,  at the core center far away from the FFLO nodal plane, we can also expect that odd parity pairing dominates. 

Figure \ref{f3} depicts the results at the FFLO nodal plane, $ (x,z)=(25\xi,0)$. The results are qualitatively similar to those in Fig. \ref{f2}. 
Quantitative difference may originate from the dimensionality of the Andreev bound states: 1D in vortex line and 2D in the FFLO nodal plane. 
The local DOS at $\mu_B H=0$ shows a zero energy peak by the presence of the odd frequency pairing for $\mu_B H=0$ as seen in Figs. \ref{f3} (b) and (c). Finite (but small) magnitude of the even frequency pairing at $\mu_B H=0$ in Fig. \ref{f3} (b) is physically due to the penetration of the even frequency pairing at the intersection point into the FFLO nodal plane. This results from the fact that the relation $\hat \Delta(x')=-\hat \Delta(-x')$ does not hold any more at the the FFLO nodal plane due to the presence of the vortex. 
With increasing $\mu_B H$, these structures are split and even frequency pairing is enhanced.

Figure \ref{f4} shows the results far away from the nodal plane and the core $(x,z)=(25\xi,25 \xi)$. The results are qualitatively similar to those in Fig. \ref{f1}. At $\mu_B H=0$, local DOS shows a gap structure with dominant even frequency component for $\mu_B H=0$ as seen in Figs. \ref{f4} (b) and (c). 
With the increase of $\mu_B H$, these structures are split and odd frequency pairing is enhanced. 

In Fig. \ref{f5}, we show (a, d, g) normalized local DOS, (b, e, h)  magnitude of even frequency pairing, and (c, f, i )  magnitude of odd frequency pairing at $\mu_B H=0$ with (a-c) $E=0$,  (d-f) $E=0.5 \Delta_0$ and (g-i) $E= \Delta_0$.
At $E=0$, the DOS shows a strong peak at the FFLO nodal plane and vortex line but at the intersection point, it is suppressed. Correspondingly, large magnitude of the odd frequency (spin singlet odd parity)  pairing is seen at the FFLO nodal plane and vortex line, while at the intersection point or far away from both the FFLO nodal plane and vortex line, only even frequency (spin singlet even parity) pairing exists. With increasing $E$, the DOS is reduced at the FFLO nodal plane and vortex line while it increases away from the FFLO nodal plane and vortex line. The structures of even and odd frequency pairings are smeared  with the increase of $E$. 

The corresponding  results at $\mu_B H=0.5 \Delta_0$ is shown in Fig. \ref{f6}. It is seen that even and odd frequency pairings are spatially distributed in a similar way. At $E=0$, the DOS has a similar structure to that in Fig. \ref{f5} (d) due to the Zeeman spliting $\mu_B H=0.5 \Delta_0$. As shown in Fig. \ref{f6} (b),  the structure of even frequency pairing is also simlar to  that in Fig. \ref{f5} (e). On the other hand, the behavior of odd frequency pairing in Fig. \ref{f6} (c) is rather different from that in  Fig. \ref{f5} (f). This is because by introducing the Zeeman spliting, spin triplet pairing, especially odd frequency  spin triplet  even parity pairing, emerges, which will be discussed together with Fig. \ref{f10} below. 
At $E=0.5 \Delta_0$, the DOS has a similar structure to that in Fig. \ref{f5} (a). As shown in Fig. \ref{f6} (f),  the structure of odd frequency pairing is also simlar to  that in Fig. \ref{f5} (c), while the even frequency pairing in Fig. \ref{f6} (e) shows a different feature from that in  Fig. \ref{f5} (b). This also results from the generation of the even frequency spin triplet odd parity pairing. At  $E= \Delta_0$, the results are similar to those in Figs. \ref{f6} (a-c).

Now, let us investigate spin and parity structure of superconducting correlation. We show  (a) magnitude of even frequency spin singlet even parity pairing, (b) magnitude of even frequency spin triplet odd parity pairing, 
(c) magnitude of odd frequency spin singlet odd parity pairing, and
(d) magnitude of odd frequency spin triplet even parity pairing, at $E=0$ and $\mu_B H=0.5 \Delta_0$ in Fig. \ref{f10}. We see that even parity pairing, $\left\langle {f_{even}^s } \right\rangle$ and $\left\langle {f_{odd}^t } \right\rangle$, is dominant at the intersection point or far away from both vortex line and FFLO nodal plane, while odd parity pairing, $\left\langle {f_{even}^t } \right\rangle$ and $\left\langle {f_{odds}^s } \right\rangle$, dominates at vortex line and FFLO nodal plane. Therefore, we find that large magnitude of odd frequency pairing away from  vortex line and FFLO nodal plane in Fig. \ref{f6} (c) is due to  the spin triplet pairing induced by the Zeeman term.  The large magnitude of even frequency pairing at vortex line and FFLO nodal plane in Fig. \ref{f6} (e) is also attributed to this effect.


\section{Vortex lattice}
\subsection{Formulation}
Here, we will consider vortex lattice in $s$-wave superconductor.
Following Ref. \onlinecite{ichiokaS}, let us explain the Eilenberger theory of vortex lattice (see also Refs. \onlinecite{Klein87,Klein89,Pottinger}). 
First, we obtain the pair potential  self-consistently 
by solving the Eilenberger equation in the Matsubara frequency. 
Next, using them, we calculate the local DOS by solving the Eilenberger equation 
in the real energy instead of the Matsubara frequency. 

In our calculation, the unit vectors of the vortex lattice are given by 
${\bf r}_1=(a_x,0)$, ${\bf r}_2=(\zeta a_x,a_y)$. 
As we consider a triangular lattice, we set $a_y/a_x=\sqrt{3}/2$ and 
$\zeta=1/2$. 
The microscopic magnetic field $\bar {\bf H}=(0,0,\bar H)$ is given by
${\bf H}({\bf r})=\nabla\times {\bf A}({\bf r}) 
=\bar {\bf H} $
where the vector potential ${\bf A}({\bf r})$ is 
${\bf A}({\bf r})=\frac{1}{2}\bar {\bf H} \times {\bf r} $
in the symmetric gauge.  

For the quasi-classical Green functions, the Eilenberger equation is given as  
\begin{widetext}
\begin{equation}
(\partial _\parallel   + i2\pi A_\parallel  /\phi _0 )a(\omega _n ,\theta ,{\bf{r}}) + \left( {2\omega _n  + \Delta ^* a(\omega _n ,\theta ,{\bf{r}})} \right)a(\omega _n ,\theta ,{\bf{r}}) - \Delta  = 0,
\label{eq:2.3}
\end{equation}
\begin{equation}
(\partial _\parallel   - i2\pi A_\parallel  /\phi _0 )b(\omega _n ,\theta ,{\bf{r}}) - \left( {2\omega _n  + \Delta b(\omega _n ,\theta ,{\bf{r}})} \right)b(\omega _n ,\theta ,{\bf{r}}) + \Delta ^*  = 0,
\label{eq:2.4}
\end{equation}
\end{widetext}
with the Matsubara frequency $\omega_n=(2n+1)\pi T$. Here, ${\bf r}$ is the center of mass  coordinate of a Cooper pair. 
The direction of the relative momentum of the Cooper pair, 
$\hat{\bf k}={\bf k}/|{\bf k}|$, is denoted by an angle $\theta$ 
measured from the $x$ axis in the hexagonal plane. Here, we define  $ \partial_\parallel = d/dr_\parallel$ and $A_\parallel=\hat{\bf k}\cdot {\bf A} 
=-\frac{1}{2}H r_\perp$.
Also, we have taken the coordinate system: 
$\hat{\bf u}=\cos\theta \hat{\bf x}+\sin\theta \hat{\bf y}$, 
$\hat{\bf v}=-\sin\theta \hat{\bf x}+\cos\theta \hat{\bf y}$, 
thus a point ${\bf r}=x \hat{\bf x}+y \hat{\bf y}$ 
is denoted as 
${\bf r}=r_\parallel \hat{\bf u}+r_\perp \hat{\bf v}$.
The first-order differential equations (\ref{eq:2.3}) and (\ref{eq:2.4}) 
are solved along the trajectory where $r_\perp$ is held constant. 
Notice that we here focus on superconductor in the type II limit, and a weak magnetic field regime so that the Zeeman term can be neglected. 

The self-consistent equation for the pair potential $\Delta({\bf r})$ 
reads
\begin{equation}
\Delta({\bf r})=V N_0 
2 \pi T \sum_{\omega_n>0} \int_0^{2\pi}{d\theta \over 2\pi}
f(\omega_n,\theta,{\bf r}) ,
\label{eq:2.7}
\end{equation}
with $f=-2ia/(1+ab)$, the density of states at the Fermi surface $N_0$, and  the pairing interaction $V$.
In our calculation, we use the relation 
\begin{equation}
\frac{1}{V N_0}=\ln \frac{T}{T_c}
+2\pi T \sum_{0<\omega_n < \omega_c} \frac{1}{|\omega_n|}, 
\label{eq:2.10}
\end{equation}
and set the energy cutoff $\omega_c=20 T_c$.

We calculate the r.h.s. of Eq. (\ref{eq:2.7}) 
using the quasi-classical Green functions obtained by Eqs. 
(\ref{eq:2.3}) and (\ref{eq:2.4}), and obtain the 
new value for $\Delta({\bf r})$. 
Using the renewed pair potential, 
we solve the Eilenberger equation (\ref{eq:2.3}) and (\ref{eq:2.4}) again. 
Using the following gap function as  an initial value, \cite{Eilenberger2}
\begin{widetext}
\begin{equation}
\Delta({\bf r})= \left(\frac{2 a_y}{a_x}\right)^{1/4}\sum_{p=-\infty}^\infty 
\exp\left\{ -\pi \frac{a_y}{a_x} \left(\frac{y+y_0}{a_y}+p \right)^2 
+2 \pi i \left[ p \left( \frac{x_0}{a_x} + \frac{\zeta}{2}p \right) 
+ \left( \frac{y_0}{a_y} + p \right) \frac{x}{a_x}  \right] \right\} 
\exp \left( i\pi \frac{xy}{a_x a_y} \right) ,
\label{eq:2.11}
\end{equation}
\end{widetext}
 we repeat this simple iteration procedure more 
than 20 times, and obtain a sufficiently self-consistent solution for 
$\Delta({\bf r})$. 
In Eq. (\ref{eq:2.11}), the r.h.s. is the Abrikosov solution of the 
vortex lattice, where we use the relation $Ha_x a_y / \phi_0 =1$. 
The factor $\exp(i\pi xy /a_x a_y)$ is due to the gauge 
transformation from the Landau gauge to the symmetric gauge. 
We set ${\bf r}_0=(x_0,y_0)= -\frac{1}{2}({\bf r}_1 + {\bf r}_2) $ 
so that one of the vortex centers locates at the origin of the coordinate. 
Note that since we determine the gap function self-consistently, our theory is applicable to any magnetic field below the upper critical field, although we adopt the Abrikosov solution as an initial value.

The physical quantities are calculated in a similar way to the previous section. The DOS is calculated as 
\begin{eqnarray}
N({\bf{r}}, E) = \int_0^{2\pi } {\frac{{d\theta }}{{2\pi }}} {\mathop{\rm Re}\nolimits} \left[ {(1 + ab)^{ - 1} (1 - ab)} \right]_{\omega _n  \to E + i\delta }.
\end{eqnarray}

Even (odd) frequency component $f_{even(odd)}$ is given by
\begin{equation}
f_{even(odd)}  = \left\{ {f(\theta ) + ( - )f(\theta  + \pi )} \right\}/2.
\end{equation}
Its average is defined as
\begin{equation}
\left\langle {f_{even(odd)} } \right\rangle ^2  = \frac{1}{{2\pi }}\int_0^{2\pi } {\left| {f_{even(odd)} } \right|^2 d\theta } .
\end{equation}
Note that since spin is conserved, \textit{all the above pairings are spin singlet}. 

In the following, we introduce $R_0$, the transition temperature $T_C$, and $H_0$ as units of length, temperature, and magnetic field, respectively, where $R_0=\hbar v_{\rm F}/2 \pi k_{\rm B} T_{\rm c}$ and $H_0=\hbar c /2|e|R_0^2$. 
Also, we fix  $\delta$ as $\delta=0.05\Delta_0$ where $\Delta_0$ is the bulk value of the gap function at $T=0$.

\subsection{Results}
Here, we consider two cases:  low field case where the distance between vortices is large, and the overlap effect of the vortex cores is weak, and  high field case as an opposite situation. In the former case, we choose  $H/H_0=0.05$ and in the latter we set $H/H_0=0.5$.
The temperature of the system is fixed as $T/T_C=0.5$.

\begin{figure*}[htb]
\begin{center}
\scalebox{0.4}{
\includegraphics[width=31.0cm,clip]{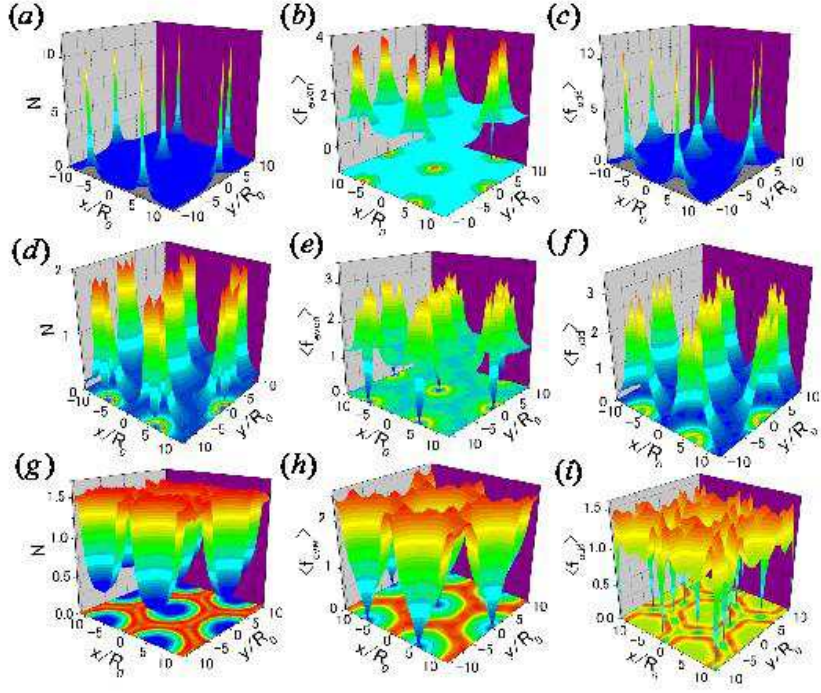}}
\end{center}
\caption{(color online) (a,d,g) normalized local DOS. (b,e,h) magnitude of even frequency pairing. (c,f,i) magnitude of odd frequency pairing. (a-c) $E=0$.  (d-f) $E=0.5 \Delta_0$.  (g-i) $E= \Delta_0$.  Here, we set $H/H_0=0.05$. } \label{f7}
\end{figure*}

\begin{figure*}[htb]
\begin{center}
\scalebox{0.4}{
\includegraphics[width=31.0cm,clip]{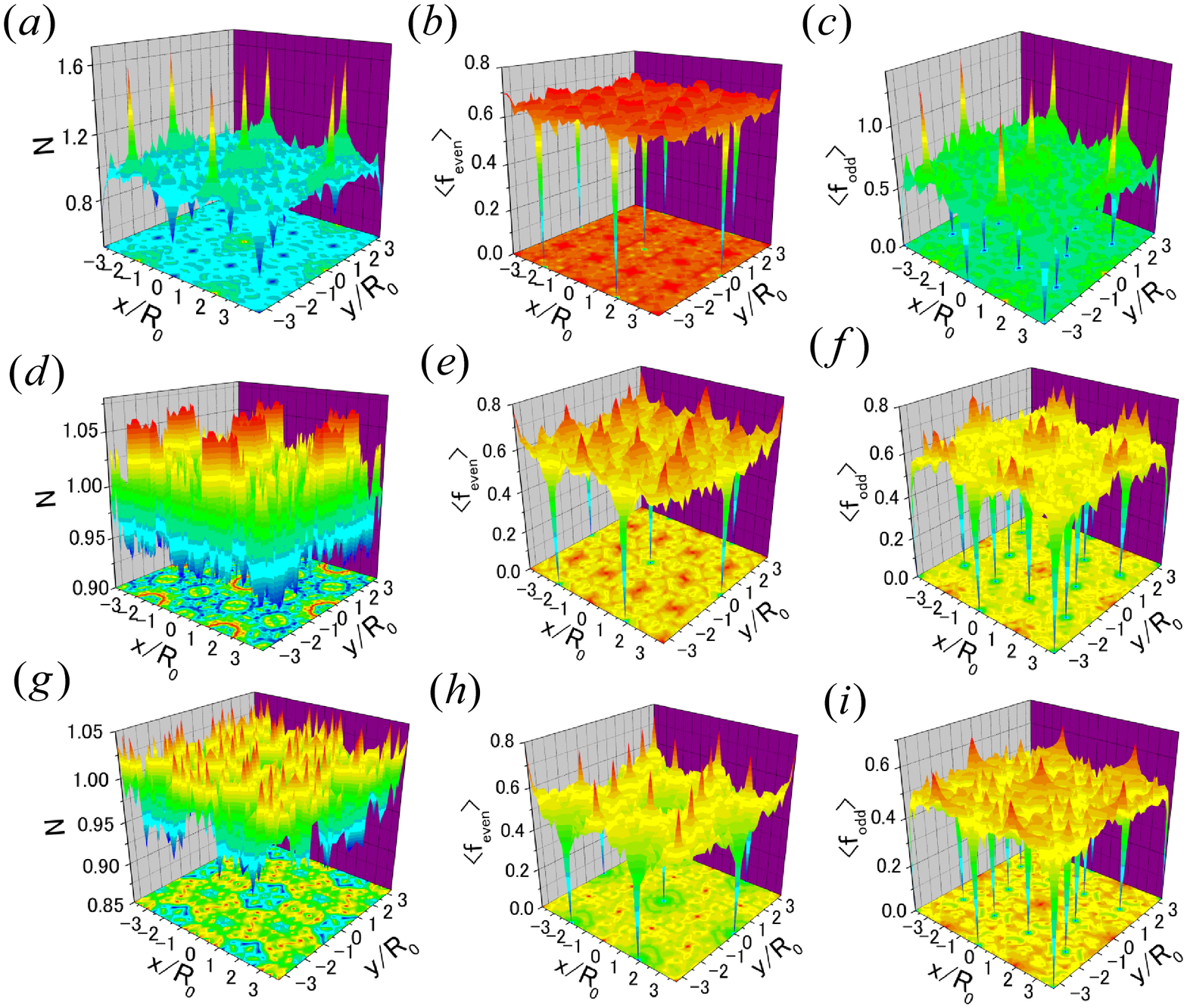}}
\end{center}
\caption{(color online) (a,d,g) normalized local DOS. (b,e,h) magnitude of even frequency pairing. (c,f,i) magnitude of odd frequency pairing. (a-c) $E=0$.  (d-f) $E=0.5 \Delta_0$.  (g-i) $E= \Delta_0$.  Here, we set $H/H_0=0.5$. } \label{f8}
\end{figure*}

\begin{figure}[htb]
\begin{center}
\scalebox{0.4}{
\includegraphics[width=22.0cm,clip]{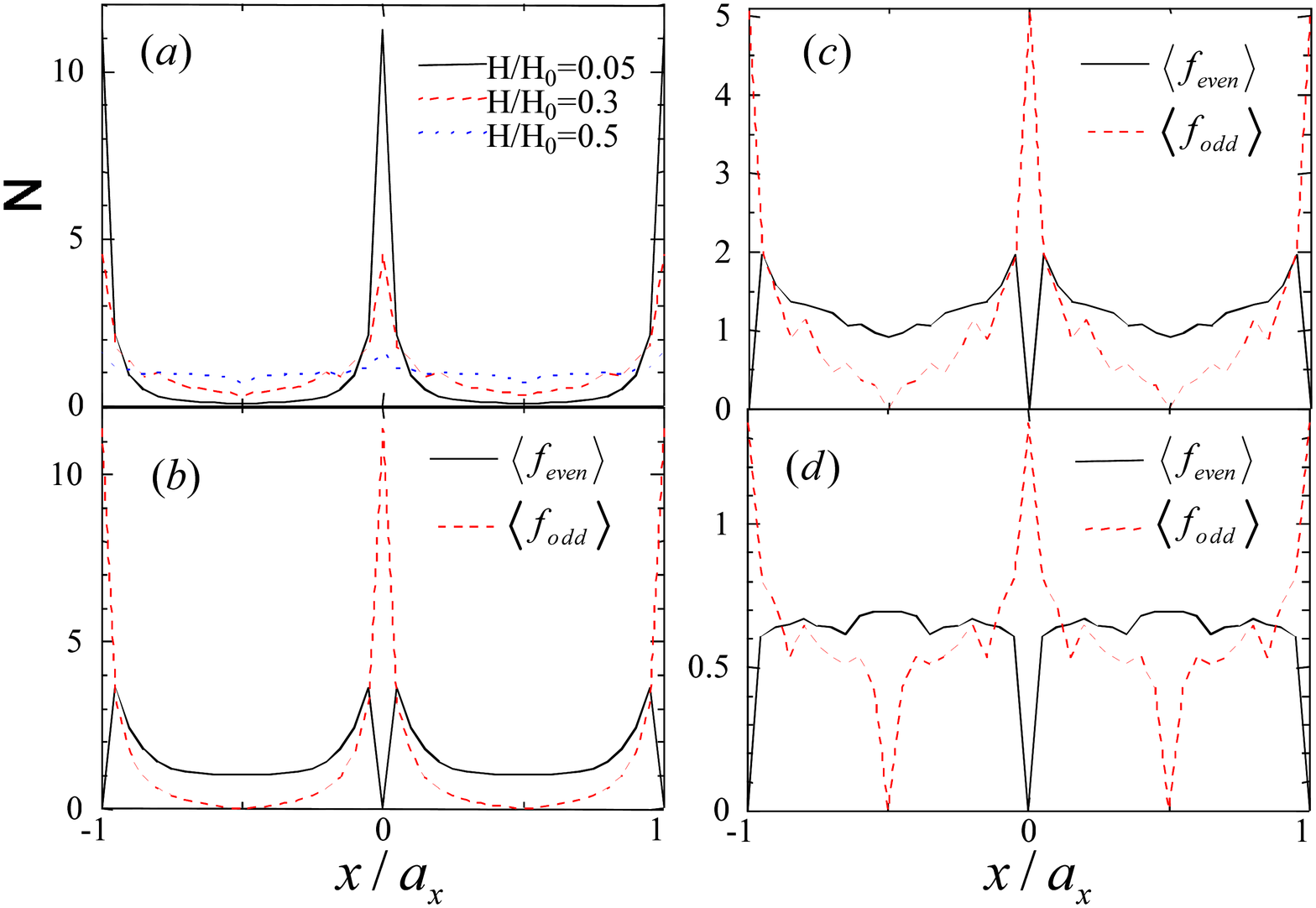}}
\end{center}
\caption{(color online) (a) normalized local DOS, and magnitude of even and frequency pairings for (b) $H/H_0$=0.05, (c) $H/H_0$=0.3, and (d) $H/H_0$=0.5 at $E=y=0$.  } \label{f9}
\end{figure}

Figure \ref{f7} shows the results at $H/H_0=0.05$. 
 At $E=0$, the Andreev bound states are seen. The odd frequency pairing shows a similar structure to the DOS, while even frequency pairing is absent at the core center although it has a large value near the core center. As $E$ increases, the DOS and the odd frequency component at the core center are reduced but they increase away from the core center. 
We find that at the core center, only odd frequency pairing is present while at the midpoint of the vortex lines, only even frequency pairing exists irrespective of the energy.
Thus, the odd and even frequency pairings also form the lattice in the vortex lattice state. In Ref. \onlinecite{Yokoyamavortex}, it is clarified that only odd frequency pairing is present at the core center of an isolated vortex. We see that this is also the case for the vortex lattice. We have also found that the orbital pairing symmetry of superconducting correlation contains not only $p$-wave component but also  higher harmonics, such as $d$-wave or $f$-wave components, in the vortex lattice since the  rotational symmetry is broken, in contrast to the single vortex case.\cite{Yokoyamavortex}

We show the results at $H/H_0=0.5$ in Fig. \ref{f8}. A qualitatively similar tendency to Fig. \ref{f7} is seen. 
However, compared to Fig. \ref{f7}, the vortex spacing is reduced and the overlap effect of the vortex cores reduces the Andreev bound states. Also, the magnitudes of the even and odd frequency components are suppressed. Meanwhile, DOS away from the core increases, and the mangitude of the odd frequency pairing becomes comparable to that of the even frequency pairing. 

To see how even and odd frequency pairings are influenced by the magnetic field in more detail, we depict the spatial dependence of (a) normalized local DOS for various $H$, and magnitude of even and frequency pairings for (b) $H/H_0$=0.05, (c) $H/H_0$=0.3, and (d) $H/H_0$=0.5 at $E=y=0$ in Fig. \ref{f9}. We see that at the core center, only odd frequency pairing exists while at the midpoint of the vortex lines, only even frequency pairing is present, which is reflected in the DOS as seen in  Fig. \ref{f9} (a). The magnitude of even and odd frequency pairings are reduced by increasing magnetic field. For higher energy, we also find a similar tendency (not shown).

\section{Summary}
In summary, we have studied pairing symmetry in the FFLO vortex and vortex lattice. 
We showed analytically 
that at the intersection point of FFLO nodal plane and vortex line, only even frequency pairing is present if the Zeeman splitting is negligibly small. With increasing Zeeman splitting, odd frequency pairing also emerges there. Therefore, the gap structure of the DOS at the intersection point predicted in Refs.~\onlinecite{mizushima2,Ichioka}  can be regarded as a manifestation of the even frequency pairing. 

In the vortex lattice, at the core centers, only odd frequency pairing is present while at the midpoint of the vortex lines, only even frequency pairing
appears irrespective of the energy. Thus, the odd and even frequency pairings also form the lattice in the vortex lattice state.

We have explained the electronic structure in vortex systems in terms of the pairing symmetry, \textit{even and odd frequency pairings}, which has been understood with the quasiparticle picture to date.
 Our approach can be extended to vortex system in other fields. 
Recently, vortex state has been realized in cold atoms. \cite{Zwierlein}
To study pairing symmetry in vortex state of cold atoms, as is done in this paper, would give an insight into vortex physics in cold atoms. 

T.Y. acknowledges support by the JSPS.

%

\end{document}